\documentclass[12pt]{amsart}

\newcommand{\R}{\mathbb R}
\newcommand{\HH}{\mathcal H}
\newcommand{\x}{\mathbf x}

\newcommand{\Symb}{Symb}
\newcommand{\nn}{\mathbf n}
\newcommand{\vol}{\mathbf{vol}}

\newcommand{\eqdef}{\stackrel{\text{def}}{=}}
\newcommand{\Id}{Id}
\newcommand{\reg}{\text{reg}}
\newcommand{\y}{\mathbf y}

\begin{document}

\title[A problem related to quantization of fields]{A new mathematical problem related to quantization of fields}
\author{A. V. Stoyanovsky}
\thanks{Partially supported by the grant RFBR 10-01-00536.}
\begin{abstract}
This paper is a survey of author's mathematical and logical study of the problem of quantization of fields.
\end{abstract}
\email{alexander.stoyanovsky@gmail.com}
\address{Russian State University of Humanities}
\maketitle
\tableofcontents

\section*{Introduction}

This paper is a survey of author's program of study of mathematical and logical problems behind quantization of fields
[1--19]. The central one of these problems is the problem of (deformation) quantization of the
Poisson algebra of observables on a space-like surface. We must warn the reader that we completely do not have
an intuitive feeling of whether this mathematical problem has a positive or negative mathematical solution
(an idea of possible approach is discussed in \S3), or it is an unsolvable problem
in the spirit of G\"odel's incompleteness theorem and Continuum Hypothesis. V.~P.~Maslov informed me that he also came
to G\"odel's theorem in his comparison of combinatorial statistical mechanics and number theoretical secondary quantization
of bosons and fermions.

The contents of the paper is the following.
First, we recall the generally covariant Hamiltonian formalism in classical field theory. We show that the widespread
opinion that the Hamiltonian field theory requires choice of time axis and hence is non-relativistically
invariant, is wrong. A detailed analysis of the multidimensional variational principle
leads to a generally covariant Hamiltonian formalism and Hamilton--Jacobi theory.

Second, we construct the Poisson algebra of classical field theory observables on a space-like surface, and pose
the problem of its quantization. We discuss how this quantization could be related to the construction of quantum field
theory $S$-matrix beyond the scope of perturbation theory.

Finally, we discuss a possible approach to the problem of finding the appropriate space of states in quantum field theory,
generalizing the space of distributions in the case of finite dimensional phase space.

I thank V.~V.~Dolotin and V.~P.~Maslov for understanding, fruitful discussions, constant help and encouragement.

\section{Classical field theory}

The idea of quantization of classical field theory is related with the idea of Feynman path integral.
This integral describes the imagined generalization of the wave theory to the situation of a multidimensional
variational problem. In order to formalize this idea, we first recall a mathematical transformation of a
variational principle which is the direct generalization of the Hamilton--Jacobi theory to multidimensional
variational problems. This is quite similar to the way of derivation of the Schrodinger equation in non-relativistic
quantum mechanics.

\subsection{Formula for variation of action}
Consider the action functional of the form
\begin{equation}
J=\int_D F(x^0,\ldots,x^n,\varphi^1,\ldots,\varphi^m,\varphi^1_{x^0},\ldots,\varphi^m_{x^n})\,
dx^0\ldots dx^n,
\end{equation}
where $x^0,\ldots,x^n$ are the space-time variables, $\varphi^1,\ldots,\varphi^m$ are the field
variables, $\varphi^i_{x^j}=\frac{\partial\varphi^i}{\partial x^j}$,
and integration goes over an $(n+1)$-dimensional surface $D$ (the graph of the functions $\varphi^i(x)$)
with the boundary $\partial D$ in the space $\R^{m+n+1}$.

Assume\, that\, for\, each\, $n$-dimensional\, parameterized\, surface\, $C$\,
in $\R^{m+n+1}$, given by the equations
\begin{equation}
x^j=x^j(s^1,\ldots,s^n),\ \ \varphi^i=\varphi^i(s^1,\ldots,s^n)
\end{equation}
and sufficiently close (in $C^\infty$-topology) to a fixed $n$-dimensional surface,
there exists a unique $(n+1)$-dimensional surface $D$ with the boundary $\partial D=C$ which is an
extremal of the integral (1), i.~e. the graph of a solution to the Euler--Lagrange equations.
Denote by $S=S(C)$ the value of the integral (1) over the surface $D$.

Then one has the following well known formula for variation of the functional $S$:
\begin{equation}
\delta S=\int_C \left(\sum\pi^i\delta\varphi^i-\sum H^j\delta x^j\right)\,ds,
\end{equation}
or
\begin{equation}
\begin{aligned}{}
\frac{\delta S}{\delta\varphi^i(s)}&=\pi^i(s),\\
\frac{\delta S}{\delta x^j(s)}&=-H^j(s),
\end{aligned}
\end{equation}
where
\begin{equation}
\begin{aligned}{}
\pi^i&=\sum_l(-1)^lF_{\varphi^i_{x^l}}
\frac{\partial(x^0,\ldots,\widehat{x^l},\ldots,x^n)}{\partial(s^1,\ldots,s^n)},\\
H^j&=\sum_{l\ne j}(-1)^lF_{\varphi^i_{x^l}}\varphi^i_{x^j}
\frac{\partial(x^0,\ldots,\widehat{x^l},\ldots,x^n)}{\partial(s^1,\ldots,s^n)}\\
&+(-1)^j(F_{\varphi^i_{x^j}}\varphi^i_{x^j}-F)
\frac{\partial(x^0,\ldots,\widehat{x^j},\ldots,x^n)}{\partial(s^1,\ldots,s^n)}.
\end{aligned}
\end{equation}

Here
$\frac{\partial(x^1,\ldots,x^n)}{\partial(s^1,\ldots,s^n)}=\left|
\frac{\partial x^j}{\partial s^i}\right|$ is the Jacobian;
the cap over a variable means that the variable is omitted; summation
over the index $i$ repeated twice is assumed. For derivation of this formula see, for example, [2], [7],
or [20].

Note that the coefficients before the Jacobians in the formula for $H^j$
coincide, up to sign, with the components of the energy-momentum tensor.
The variables $\pi^i(s)$ are called {\it canonically conjugate} to the variables $\varphi^i(s)$.

Note also that the quantities $\pi^i$ and $H^j$ depend on the numbers $\varphi^i_{x^j}$
characterizing the tangent plane to the $(n+1)$-dimensional surface $D$.
These numbers are related by the system of equations
\begin{equation}
\varphi^i_{x^j}x^j_{s^k}=\varphi^i_{s^k},\ \ \ i=1,\ldots,m,\ k=1,\ldots,n.
\end{equation}
Hence, only $m(n+1)-mn=m$ numbers among $\varphi^i_{x^j}$ are independent. Therefore
$m+n+1$ quantities $\pi^i$ and $H^j$ are related, in general, by $n+1$ equations. $n$ of these equations
are easy to find:
\begin{equation}
\pi^i\varphi^i_{s^k}-H^jx^j_{s^k}=0,\ \ \ k=1,\ldots,n.
\end{equation}
The remaining $(n+1)$-th equation depends on the form of the function $F$. Denote it by
\begin{equation}
\HH(x^j(s),\varphi^i(s),x^j_{s^k},\varphi^i_{s^k},\pi^i(s),-H^j(s))=0.
\end{equation}

From $n+1$ equations (7) and (8) one can, in general, express the quantities $H^j$
as functions of $\pi^i$ (and of $x^l$, $u^i$, $x^l_{s^k}$, $\varphi^i_{s^k}$):
\begin{equation}
H^j=H^j(x^l,\varphi^i,x^l_{s^k},\varphi^i_{s^k},\pi^i),\ \ \ j=0,\ldots,n.
\end{equation}

\subsection{The\,\, generalized\,\, Hamilton--Jacobi\,\, equation}\,\,
Substituting (4) into equations (7,8) or into equations (9), we obtain
\begin{equation}
\begin{aligned}{}
\frac{\delta S}{\delta\varphi^i(s)}\varphi^i_{s^k}+\frac{\delta S}{\delta x^j(s)}
x^j_{s^k}&=0,\ \ \ k=1,\ldots,n,\\
\HH\left(x^j,\varphi^i,x^j_{s^k},\varphi^i_{s^k},
\frac{\delta S}{\delta\varphi^i(s)},\frac{\delta S}{\delta x^j(s)}\right)&=0,
\end{aligned}
\end{equation}
or
\begin{equation}
\frac{\delta S}{\delta x^j(s)}+H^j\left(x^l,\varphi^i,x^l_{s^k},\varphi^i_{s^k},
\frac{\delta S}{\delta\varphi^i(s)}\right)=0,\ \ \ j=0,\ldots,n.
\end{equation}
The system of equations (10) or (11), relating the values of variational derivatives
of the functional $S$
at one and the same point $s$, can be naturally called the generalized Hamilton--Jacobi
equation. The first $n$ equations of the system (10) correspond to the fact that the function $S$ does
not depend on concrete parameterization of the surface $C$.

\medskip
{\bf Example (scalar field with self-action).}
Let
\begin{equation}
F(x^\mu,\varphi,\varphi_{x^\mu})=\frac{1}{2}(\varphi_{x^0}^2-\sum_{j\ne0}\varphi_{x^j}^2)-V(x,\varphi)
=\frac12\varphi_{x^\mu}\varphi_{x_\mu}-V(x,\varphi)
\end{equation}
in the standard relativistic notations, where the index $\mu$ is pushed down using the metric
$(dx^0)^2-\sum_{j\ne0}(dx^j)^2$.
A computation gives the following generalized Hamilton--Jacobi equation:
\begin{equation}
\begin{aligned}{}
x^\mu_{s^k}\frac{\delta S}{\delta x^\mu(s)}+
\varphi_{s^k}\frac{\delta S}{\delta\varphi(s)}&=0,\ \ k=1,\ldots,n,\\
\vol\frac{\delta S}{\delta\nn(s)}+
\frac{1}{2}\left(\frac{\delta S}{\delta\varphi(s)}\right)^2
+\frac12\vol^2&d\varphi(s)^2+\vol^2V(x(s),\varphi(s))=0.
\end{aligned}
\end{equation}
Here $\vol^2=D^\mu D_\mu$ is the square of the volume element on the surface,
$D^\mu=(-1)^\mu \frac{\partial(x^0,\ldots,\widehat{x^\mu},\ldots,x^n)}{\partial(s^1,\ldots,s^n)}$,
the vector $(D_\mu)=\vol\cdot\nn$ is proportional to the unit normal $\nn$ to the surface,
the number $\vol\frac{\delta S}{\delta\nn(s)}=D_\mu\frac{\delta S}{\delta x^\mu(s)}$ is proportional to
the variation $\frac{\delta S}{\delta\nn(s)}$ of the functional $S$ under the change of the surface in the
normal direction, and the number
$\vol^2d\varphi(s)^2=(D_\mu\varphi_{x^\mu})^2-(\varphi_{x^\mu}\varphi_{x_\mu})(D_\nu D^\nu)$ is proportional
to the scalar square
$d\varphi(s)^2$ of the differential $d\varphi(s)$ of the function $\varphi(s)$ on the surface.
\medskip

The generalized Hamilton--Jacobi equation was written in particular cases by many authors,
see, for example, the book [21] and references therein. In [21] one can also find
a theory of integration of the generalized Hamilton--Jacobi equation in the particular case of
two dimensional variational problems but in somewhat non-convenient notations, and in [2,7] a theory in the general case.

\subsection{Generalized canonical Hamilton equations}
Suppose that the surface $D$ is parameterized by the coordinates
$s_1$, $\ldots$, $s_n$, $t$. The generalized canonical Hamilton equations express the dependence
of the variables $\pi^i,\varphi^i$ on $t$,
if we assume that the dependence of $x^j$ on $(s,t)$ is given. The equations read
\begin{equation}
\begin{aligned}{}
\varphi^i_t&=\frac{\delta}{\delta\pi^i(s)}\int H^jx^j_t(s')\,ds', \\
\pi^i_t&=-\frac{\delta}{\delta\varphi^i(s)}\int H^jx^j_t(s')\,ds'.
\end{aligned}
\end{equation}
For their derivation, see [2,7]. They are equivalent to the Euler--Lagrange equations. They can be also written in the
following form:
\begin{equation}
\frac{\delta\Phi(\pi^i(\cdot),\varphi^i(\cdot);x^j(\cdot))}{\delta x^j(s)}=\{H^j(s),\Phi\},
\end{equation}
where $\Phi(\pi^i(\cdot),\varphi^i(\cdot);x^j(\cdot))$ is an arbitrary functional of functions $\varphi^i(s)$, $\pi^i(s)$
changing together with the surface $x^j=x^j(s)$, and
\begin{equation}
\{\Phi_1,\Phi_2\}=\sum_i\int\left(\frac{\delta\Phi_1}{\delta \pi^i(s)}
\frac{\delta\Phi_2}{\delta\varphi^i(s)}-\frac{\delta \Phi_1}{\delta\varphi^i(s)}
\frac{\delta\Phi_2}{\delta\pi^i(s)}\right)ds
\end{equation}
is the Poisson bracket of two functionals $\Phi_l(\pi^i(\cdot),\varphi^i(\cdot))$, $l=1,2$. In [2,7]
the generalized canonical Hamilton equations are identified with the equations of characteristics
for the generalized Hamilton--Jacobi equation.

In one-dimensional variational calculus, the Hamilton--Jacobi theory is the most powerful of the known methods of
integration of the canonical Hamilton equations. It would be interesting to check whether this holds for
multidimensional variational principles, by integrating the generalized Hamilton--Jacobi equation
(finding a full integral [2,7]) for,
say, the Einstein equation or the Yang--Mills equation; this would lead to a new way of integration of
Einstein or Yang--Mills equations.

\section{The Poisson algebra of classical field theory observables on a space-like surface
and the problem of its quantization}

\subsection{Construction of the Poisson algebra of classical Hamiltonians in field theory}

Let us pose the following question. Let $\varphi=\varphi(\x)$ be the
classical (scalar for simplicity) field on a space-like surface $C$ in space-time $\R^{n+1}$,
$\x=(x_1,\ldots,x_n)$ be coordinates on the surface, and let $\pi=\pi(\x)$ be the conjugate momentum.
Both $\varphi$ and $\pi$ belong to the Schwartz space $S$ of smooth functions rapidly decreasing at infinity.
The question is: what functionals $H(\varphi,\pi)$ can be classical field theory Hamiltonians? This class of functionals
should contain the known examples, and should agree with the picture for free field (quadratic Hamiltonians).

Recall that the usual known examples of Hamiltonians are integrals over $C$ of polynomial densities depending locally
on $\varphi$ and $\pi$. On the other hand, in the case of free Klein--Gordon field, the evolution operators from
one space-like surface to another belong to the group of continuous symplectic transformations of the
symplectic topological vector space $S\oplus S$. Therefore it is natural to expect that the quadratic Hamiltonians
form the Lie algebra of this group.

It is easy to see that the answer to our question should be the following.
\medskip

{\bf Definition.}[17] A continuous polynomial functional
$H(\varphi,\pi)$ on $S\times S$ is called a Hamiltonian (or a symbol) if its first
functional differential $\delta H$, which is a linear functional
on test functions $(\delta\varphi,\delta\pi)\in S\oplus S$ for every $(\varphi,\pi)$, belongs to $S\oplus S\subset
S'\oplus S'$ (here $S'$ is the space of tempered distributions dual to $S$). Moreover, $\delta H$ should be
infinitely differentiable as an $S\oplus S$-valued functional on $S\times S$.

Denote the space of symbols by $\Symb$.
\medskip

{\bf Proposition.} $\Symb$ is a topological Poisson algebra with respect to the
standard Poisson bracket
\begin{equation}
\{H_1,H_2\}=\int\left(\frac{\delta H_1}{\delta\pi(\x)}\frac{\delta H_2}{\delta\varphi(\x)}
-\frac{\delta H_1}{\delta\varphi(\x)}\frac{\delta H_2}{\delta\pi(\x)}\right)d\x.
\end{equation}
\medskip

{\it Proof} is straightforward.
\medskip

Note that, unlike any differentiable functional on the space $S\times S$,
any $H\in\Symb$ generates a well defined Hamiltonian flow on the phase space $S\times S$.

Any $H\in\Symb$ has the form
\begin{equation}
\begin{aligned}{}
H(\varphi,\pi)&=\sum_{k=0}^N\sum_{l=0}^M H_{k,l}(\varphi,\pi),\\
H_{k,l}(\varphi,\pi)&=\frac1{k!l!}\int a_{k,l}(\x_1,\ldots,\x_k;\y_1,\ldots,\y_l)\\
&\times\varphi(\x_1)\ldots\varphi(\x_k)\pi(\y_1)\ldots\pi(\y_l)d\x_1\ldots d\x_k d\y_1\ldots d\y_l
\end{aligned}
\end{equation}
for certain tempered distributions $a_{k,l}$ symmetric in $\x_1,\ldots,\x_k$ and in $\y_1,\ldots,\y_l$ (by the Schwartz
kernel theorem).
\medskip

{\bf Definition.} If all $a_{k,l}$ are smooth functions rapidly decreasing at infinity then the Hamiltonian
$H$ is called {\it regular}. Otherwise it is called {\it singular}.
\medskip

Denote the subspace of $\Symb$ consisting of regular Hamiltonians by $\Symb^{\reg}$. Clearly,
it is a Poisson subalgebra dense in $\Symb$.

\subsection{The quantization problem and renormalization on space-like surfaces}

\subsubsection{Definition.}\, A\, {\it quantization}\, ({\it deformation\, quantization})\,
of\, the Poisson algebra $\Symb$ is a continuous associative product
\begin{equation}
(H_1,H_2)\to H_1*H_2
\end{equation}
on the topological vector space $\Symb$ smoothly (resp. formally) depending on a parameter $h$ such that
\begin{equation}
\begin{aligned}{}
\text{ i) }& H_1*H_2=H_1H_2+O(h),\\
\text{ii) }& [H_1,H_2]\eqdef H_1*H_2-H_2*H_1=ih\{H_1,H_2\}+O(h^2),\\
\text{ iii) }& \text{if }H_1,H_2\text{ are quadratic }(k+l\le 2\text{ in (18)})\text{ then }\\
&[H_1,H_2]=ih\{H_1,H_2\}.
\end{aligned}
\end{equation}
Two (deformation) quantizations $*_1$ and $*_2$ are called {\it equivalent} if there exists a linear map
from $\Symb$ to itself smoothly (resp. formally) depending on $h$, of the form $\Id+O(h)$, which takes
$*_1$ to $*_2$.
\medskip

{\bf Problems.} Find a (deformation) quantization of $\Symb$. Classify all (deformation) quantizations
up to equivalence.

\subsubsection{Discussion\, of\, the\, quantization\, problem\, and\, renormalization}
Clearly, the Poisson subalgebra
$\Symb^\reg\subset\Symb$ admits a quantization to the algebra of regular functional differential operators
with the usual formula for the product of symbols,
\begin{equation}
\begin{aligned}{}
H_1*_{Diff}H_2(\varphi,\pi)=&\exp\left(ih\int\frac{\delta}{\delta\pi_1(\x)}\frac{\delta}{\delta\varphi_2(\x)}d\x\right)\\
&H_1(\varphi_1,\pi_1)H_2(\varphi_2,\pi_2)|_{\varphi_1=\varphi_2=\varphi,\pi_1=\pi_2=\pi},
\end{aligned}
\end{equation}
or to the Moyal algebra
\begin{equation}
\begin{aligned}{}
H_1*_{Moyal}H_2(\varphi,\pi)=&\exp\frac{ih}2\int\left(\frac{\delta}{\delta\pi_1(\x)}\frac{\delta}{\delta\varphi_2(\x)}-
\frac{\delta}{\delta\pi_2(\x)}\frac{\delta}{\delta\varphi_1(\x)}\right)d\x\\
&H_1(\varphi_1,\pi_1)H_2(\varphi_2,\pi_2)|_{\varphi_1=\varphi_2=\varphi,\pi_1=\pi_2=\pi},
\end{aligned}
\end{equation}
or to the Wick algebra
\begin{equation}
\begin{aligned}{}
H_1*_{Wick}H_2(\varphi,\pi)=&\exp\left(h\int\frac{\delta}{\delta\varphi^-_1(\x)}
\frac{\delta}{\delta\varphi^+_2(\x)}d\x\right)\\
&H_1(\varphi^+_1,\varphi^-_1)H_2(\varphi^+_2,\varphi^-_2)|_{\varphi_1=\varphi_2=\varphi,\pi_1=\pi_2=\pi}.
\end{aligned}
\end{equation}
(Cf. [22].)
It is easy to see that these three quantizations are equivalent.

However, these quantizations cannot be extended to the whole $\Symb$, because the products contain pairings of
higher functional derivatives which are distributions rather than smooth functions. (Recall that in $\Symb$
only the first functional derivatives are functions.) Therefore, a product formula for $\Symb$ should be
constructed in a different way.

Assume that such a quantization $*$ has been found. Moreover, assume that we have an inclusion of quantum algebras
\begin{equation}
R:(\Symb^\reg, *_{Wick})\hookrightarrow(\Symb,*).
\end{equation}
Note that for $H\in\Symb^\reg$, we have, in general, $R(H)\ne H$. Moreover,
for a family $H_\Lambda$, $\Lambda\in\R$ of regular
Hamiltonians which tend to a non-regular Hamiltonian $H$ as $\Lambda\to\infty$, we have $R^{-1}(H_\Lambda)\to\infty$.

We call the map $H_\Lambda\to R^{-1}(H_\Lambda)$ the {\it renormalization map on space-like surfaces}.

\subsection{Relation with $S$-matrix}

Assume that a quantization of $\Symb$ has been found. Then the
element
\begin{equation}
S=T\exp\frac1{ih}\int_{-T_0}^{T_0} H(t)dt,
\end{equation}
where
\begin{equation}
\begin{aligned}{}
H(t)&=\int\frac12\left(\pi(\x)^2+\sum_{j=1}^n\left(\frac{\partial\varphi}{\partial x_j}\right)^2(\x)
+m^2\varphi(\x)^2\right)d\x\\
&+\int\left(\frac1{k!}g(t,\x)\varphi(\x)^k+j(t,\x)\varphi(\x)\right)d\x\in\Symb
\end{aligned}
\end{equation}
is the Hamiltonian (and the interval $(-T_0,T_0)$ contains the support of the interaction cutoff function $g(t,\x)$
and the source function $j(t,\x)$),
is a natural candidate for the role
of the (non-perturbative) Bogoliubov $S$-matrix [23]. Indeed, relativistic considerations of \S1 show that it satisfies
Bogoliubov's principles of causality, unitarity, Lorentz invariance, and the correspondence principle (see [23]).
The main difficulty here would be to identify the element $S$ with an operator in the Fock space, see below
for a discussion of this issue.

\section{Discussion of the problem of states}

One of the most perspective approaches to the quantization problem from \S2 would be to define the {\it space of states}
of quantum field theory on a space-like surface, so that the quantum algebra of observables would be an algebra of
operators on this space. Of course, this space cannot coincide with the Fock space, since the symplectic Lie algebra
of the space $S\oplus S$, which is a subalgebra of the algebra of quantum observables (see \S2),
does not act on the Fock space, as is well known from the work of Shale [24] and Berezin [25], cf. [26].
It is natural to expect that the space of states would contain a Hilbert space isomorphic to Fock space. The space of
states could be the infinite dimensional analog of the space of distributions, while the algebra of quantum observables
could be the infinite dimensional analog of the algebra of differential operators.

Let us propose an idea of construction of an infinite dimensional analog of the space of distributions. Let $u(x)$,
$x=(x^1,\ldots,x^N)$, be a tempered distribution. Consider the asymptotics of the integral
\begin{equation}
\int\psi(x,\varepsilon)u(x)\,dx,\,\,\varepsilon\to 0,
\end{equation}
where
\begin{equation}
\begin{aligned}{}
\psi(x,\varepsilon)
&=\exp\left\{\frac{2\pi i}\varepsilon\left(\frac12(x-x_0)^TZ(x-x_0)+p_0^T(x-x_0\right)\right\}\times\\
&\times P\left(\frac{x-x_0}{\sqrt\varepsilon}\right)
\end{aligned}
\end{equation}
is a {\it Gaussian wave packet}; here $(x_0,p_0)$ is a point of the phase space,
the symbol $T$ denotes transposing, $Z$ is a
symmetric complex $N\times N$-matrix with positive definite imaginary part,
and $P$ is a polynomial in $N$ variables. These Gaussian wave packets and their role in physical
applications have been studied in [27,28].

{\bf Problem.} Describe the set of asymptotics (27) for all tempered distributions $u$.

{\bf Question.} Can one reconstruct uniquely the distribution $u$ from the set of asymptotics (27)?

Note that Hormander [29] defined the wave front of a distribution by pairing it with wave packets with compact support,
so that the questions under discussion can be considered as development of Hormander's theory.

If the answer to the last question is ``yes'', then one could try to construct the infinite dimensional analog of the
space of distributions generalizing their ``microlocal description'' given by (27).

\end{document}